\documentclass[%
 reprint,
 amsmath,amssymb,
 aps,
]{revtex4-2}

\usepackage{graphicx}% Include figure files
\usepackage{dcolumn}% Align table columns on decimal point
\usepackage{bm}% bold math
\begin{document}

\preprint{APS/123-QED}

\title{Proposal for Superconducting Quantum Networks Using Multi-Octave Transduction to Lower Frequencies}% Force line breaks with \\

\author{Takuma Makihara}
 \email{makihara@stanford.edu}
 % \altaffiliation[Also at ]{Department of Applied Physics, Stanford University, Stanford, CA, }%Lines break automatically or can be forced with \\
\author{Wentao Jiang}%
\author{Amir H. Safavi-Naeini}%
 \email{safavi@stanford.edu}
\affiliation{%
 Department of Applied Physics, Stanford University, Stanford, CA 94305
}%

% \collaboration{MUSO Collaboration}%\noaffiliation

%  \homepage{http://www.Second.institution.edu/~Charlie.Author}
% \affiliation{
%  Second institution and/or address\\
%  This line break forced% with \\
% }%
% \affiliation{
%  Third institution, the second for Charlie Author
% \affiliation{%
%  Authors' institution and/or address\\
%  This line break forced with \textbackslash\textbackslash
% }%

% \collaboration{CLEO Collaboration}%\noaffiliation

\date{\today}% It is always \today, today,

\begin{abstract}
We propose networking superconducting quantum circuits by transducing their excitations (typically 4-8 GHz) to 100-500 MHz photons for transmission via cryogenic coaxial cables. Counter-intuitively, this frequency downconversion reduces noise and transmission losses. We introduce a multi-octave asymmetrically threaded SQUID circuit (MOATS) capable of the required efficient, high-rate transduction. For a 100-meter cable with $Q_i = 10^5$ at 10 mK, our approach achieves single-photon fidelities of 0.962 at 200 MHz versus 0.772 at 8 GHz, and triples the lower bound on quantum channel capacity. This method enables kilometer-scale quantum links while maintaining high fidelities, combining improved performance with the practical advantages of flexible, compact coaxial cables.
\end{abstract}

\maketitle

%\tableofcontents

%%%%%%%%%%%%%%%%%%%% INTRODUCTION %%%%%%%%%%%%%%%%%%%%
\section{Introduction}
Superconducting quantum circuits are a leading platform for building large-scale quantum computers~\cite{devoret2013superconducting,arute2019quantum,wu2021strong,jurcevic2021demonstration}. However, scaling up the number of qubits is a significant engineering challenge. Simply packing more qubits into a smaller footprint leads to challenges such as crosstalk~\cite{wenner2011wirebond,noroozian2012crosstalk}, losses from interfaces~\cite{geerlings2012improving,wang2015surface}, increased sensitivity to heat generation, and so on. This challenge has motivated a modular approach to building superconducting quantum computers, allowing one to focus on optimizing each module's coherence before final assembly~\cite{chou2018deterministic,gold2021entanglement,zhou2021modular,niu2023low,zhong2021deterministic,leung2019deterministic,kurpiers2018deterministic}. Nonetheless, the sheer number of physical qubits needed to perform interesting computations, along with the control wiring complexity, and the realities of current dilution fridge technology~\cite{krinner2019engineering}, make it attractive to distribute computation across several independent cryogenic units that can be independently optimized, tuned-up, and temperature-cycled. Thus, networking nearby dilution refrigerators would further increase the number of available qubits with an additional layer of modularity~\cite{brecht2016multilayer,bravyi2022future}. Quantum microwave-to-optical transducers aim to entangle qubits in different fridges using optical fibers~\cite{han2021microwave}, which can preserve quantum information over large distances at room temperature~\cite{chen2021twin,hensen2015loophole}. However, state-of-the-art microwave-to-optical transducers still suffer from poor efficiencies and have yet to be used to directly entangle qubits. Moreover, the long distances enabled by optical networking are not of primary interest when realizing a quantum computer, which may initially be located inside a single facility.

For shorter links, it is possible to directly connect different dilution fridges with cryogenic connections to build microwave and millimeter-wave quantum networks~\cite{xiang2017intracity,pechal2017millimeter}. Recent experiments demonstrated that qubits in different fridges can be entangled with high fidelity by directly coupling them through low-loss microwave waveguides maintained at 20 mK~\cite{magnard2020microwave}, and have led to loop-hole free violation of Bell's inequality~\cite{storz2023loophole}. However, microwave waveguides are challenging to scale to realize the needed channel capacities because their rigid structure and large size per channel. Superconducting coaxial cables have a number of technological advantages as compared to rectangular microwave waveguides. Their mechanical flexibility makes them easier to deploy, and their narrow diameter allows for orders of magnitude more communication channels per unit area. Such cables have already been used to show high-fidelity quantum channels within a single fridge~\cite{niu2023low,zhong2021deterministic,leung2019deterministic,kurpiers2018deterministic}. Unfortunately, their losses at microwave frequencies make them unattractive for longer links.

An intriguing alternative to both the up-conversion and direct microwave connection approaches is to convert the quantum information to \emph{lower} frequencies, closer to the baseband, before transmitting over superconducting coaxial cables. At first glance, this approach seems counter-intuitive as the number of thermal excitations increases as we lower the electromagnetic frequency. However, at lower frequencies the wavelength $\lambda$ of photons in the cable also increases, and so for a given, possibly frequency-independent cable quality factor ($Q$), the field attenuation constant $\alpha\approx \pi (Q\lambda)^{-1}$ decreases linearly with frequency, causing an exponential reduction in the attenuation $\eta_\text{cable}=e^{-2\alpha L}$ experienced in a cable of length $L$. For the small levels of attenuation needed to realize quantum links, we find that $\eta_\text{cable}\approx 1-2\alpha L$. The chance of a photon jump event, or errors occurring, during propagation through the cable is approximately $p_e = \alpha L (4\bar{n}+1)$, where $\bar{n}$ is the thermal occupation of the bath. In the high temperature limit $\bar{n}\approx kT\lambda/ch$ and we see that dependence of $p_e$ on wavelength cancels. However, at low temperatures, where $\bar{n}\ll 1$, the chance of an error in the link \emph{decreases} at longer wavelengths. Thus we see that there is always an advantage to operating at lower frequencies, and that moving to lower frequencies makes the system sufficiently robust to dissipation and noise to allow the use of superconducting coaxial cables to transmit quantum information. Moreover, measurements suggest that cable-mode $Q$s in fact increase at lower frequencies~\cite{kurpiers2017characterizing}.

In this paper, we consider transduction of microwave quantum information to lower frequencies, down to where the wavelength approaches the fridge-to-fridge separation. While some superconducting qubit systems, such as those using fluxonium circuits, natively operate at low frequencies~\cite{manucharyan2009fluxonium,nguyen2019high,zhang2021universal}, most current systems use qubits such as transmons that operate at higher frequencies ($>4$ gigahertz). For networking these higher frequency systems, a multi-octave quantum frequency converter is needed to bridge the gap from gigahertz down to hundreds of megahertz.

Previous works have demonstrated circuits capable of achieving such multi-octave frequency conversion by longitudinally coupling low-frequency resonators with frequencies of several hundred megahertz and high-frequency resonators with frequencies of several gigahertz~\cite{eichler2018realizing,bothner2021photon,rodrigues2021cooling}. However, these circuits rely on the current flowing in the low-frequency resonator modulating the external magnetic flux through a SQUID loop in the high-frequency resonator. Therefore, there is a direct trade off between large coupling rates and flux sensitivity, limiting their performance to low rates. Alternatively, optomechanical systems~\cite{aspelmeyer2014cavity} provide an attractive platform for strongly coupling excitations from two very different frequency bands due to the radiation pressure nonlinearity. However we require our low-frequency mode to strongly couple to a coaxial cable, which is challenging for mechanical resonators. 

We therefore propose using an asymmetrically threaded SQUID (ATS) circuit~\cite{lescanne2020exponential}. These circuits, much like the SNAIL circuits~\cite{frattini20173}, provide a fairly clean three-wave-mixing type interaction by operating at a bias-point that cancels out the four-wave-mixing or Kerr nonlinearity. Furthermore, the ATS bias-point is a flux saddle-point. Unfortunately, current approaches to using the ATS are based on weak hybridization between the two modes of interest. They are therefore not suitable for our application, as the coupling rates decrease with increasing frequency difference between the two modes, and we are interested in obtaining an extreme (``multi-octave'') frequency difference. To address this issue, we propose a modification where the ATS nonlinearly couples two electromagnetic modes of very different frequency with an essentially detuning-independent rate. We call this design a \emph{Multi-Octave ATS} (MOATS) transducer as it enables efficient coupling between electromagnetic modes with widely separated frequencies. We evaluate the performance of a simple network that uses the MOATS transducer using two key figures of merit: (1) the fidelity of transferring a single photon with Gaussian temporal mode profile, and (2) the quantum channel capacity.

Our results demonstrate that, with reasonable assumptions on cable and transducer quality factors, transducing information at 200 MHz using the MOATS leads to single-photon conversion fidelities $F\approx 0.96$, whereas transmitting the information directly at 8 GHz would lead to $F\approx 0.772$. Similarly, based on the best-known upper and lower bounds for Gaussian thermal loss quantum channel capacities~\cite{wang2022quantum}, we show that the \textit{lower bound} for channel capacity at 200 MHz is roughly three times higher than the \textit{upper bound} at 8 GHz. Finally, we propose several concrete designs for the MOATS and discuss fabrication constraints in developing these circuits.

%%%%%%%%%%%%%%%%%%%% PROPOSED SYSTEM %%%%%%%%%%%%%%%%%%%%
\section{Proposed System}
Our proposed network is based on superconducting coaxial cables. For concreteness, we consider NbTi coaxial cables that are already available at lengths over several meters~\cite{kurpiers2017characterizing}. The (field) attenuation constant $\alpha(f)$ is given by:
\begin{equation}
    \alpha(f) = \frac{\pi \sqrt{\epsilon_r}}{Q_\mathrm{i} c} f
\end{equation}
where $\epsilon_r$ is the permittivity of the coax dielectric, $Q_\mathrm{i}$ is the internal quality factor of cable standing modes, $c$ is the speed of light in vacuum, and $f$ is frequency. Previous experiments have shown that $Q_\mathrm{i}\simeq$ $10^5$ is possible at $\simeq~2$~GHz, and suggest that $Q_\mathrm{i}$ should increase at lower frequencies~\cite{kurpiers2017characterizing}. More recently, $Q_\mathrm{i}\simeq 10^6$ has been demonstrated in Al coaxial cables~\cite{niu2023low}. With these performance metrics, the attenuation constant at around 400 MHz (0.048 dB/km) is smaller than that of optical fiber at telecom frequency (0.15 dB/km).

%However, to access these low frequency cable modes, we require transducing down from typical qubit frequencies ($\simeq$ 2-8 GHz) down to below 1 GHz. Many transduction schemes that aim to bridge large frequency gaps rely on three-wave mixing, where two photons with vastly different energies can interact through a semi-classical drive field. For example, optomechanical systems provide one of the cleanest three-wave mixing interactions and can connect optical and microwave frequencies~\cite{aspelmeyer2014cavity}. However, our proposal crucially relies on strong and direct coupling of the transducer's low frequency mode to a cable, which is challenging with mechanical motion. Therefore, we seek an electromagnetic low-frequency mode. 

To access these low frequency bands, we consider three-wave mixing elements that couple electromagnetic resonances of different frequencies. Two such circuits are the so-called SNAIL~\cite{frattini20173} and ATS ~\cite{lescanne2020exponential}. We propose to use a modified ATS circuit, where the ATS inductor is shared between a high- and low-frequency electromagnetic mode, thereby giving significant zero-point fluctuations across the junctions from both the modes of our circuit. A similar approach has been previously demonstrated in generating a higher order coupling between photons in a superconducting circuit and a qubit~\cite{gely2019observation}. Our proposed circuit is shown in Figure 1a. Our designs typically have $C_a \gg C_b$ to realize an $a$-mode with a lower frequency than the $b$-mode. We can intuitively understand the modes of the circuit and their respective current distributions by considering the high and low frequency limits. At high frequencies, $C_a$ can be thought of as a short to ground, and thus the inductors $L_a$ and $L_b$ add in parallel, approximately leading to a current distribution shown in blue in Figure 1b. At low frequencies, the two inductors $L_a$ and $L_b$ add in series, approximately leading to a current distribution shown in red in Figure 1b. The capacitive coupling to the input port acts as an approximate high-pass filter, and the inductive coupling to the transmission channel, \textit{e.g.}, an NbTi superconducting coaxial cable, acts as an approximate low-pass filter. 

When the ATS is DC flux-biased to the saddle point~\cite{lescanne2020exponential}, corresponding to when an external flux of $0$ and $\pi$ thread either loop, the nonlinearity from the junctions vanish, leaving the $a$-mode and $b$-mode decoupled. Three-wave mixing between these two modes is then mediated by an RF flux pump $\epsilon_p(t)$ that threads both loops in phase. To first order in $\epsilon_p(t)$, the Hamiltonian is given by:
\begin{equation}
    \hat{H} = \hbar\omega_a \hat{a}^\dagger \hat{a} + \hbar\omega_b \hat{b}^\dagger \hat{b} - 2E_J\epsilon_p(t)\sin(\hat{\varphi}_a + \hat{\varphi}_b)
\end{equation}
where $\omega_{a(b)}$ is the frequency of the low-frequency $a$-mode (high-frequency $b$-mode), $E_J$ is the single junction energy of the ATS, and $\varphi_{a(b)}$ is the zero-point fluctuations across the junctions coming from the $a$-mode ($b$-mode). We can engineer a beam-splitter  Hamiltonian between the $a$-mode and $b$-mode by appropriate driving.  Notice that when the Hamiltonian is expanded to third order, parametrically pumping the flux line at $\omega_p = 2\omega_b - \omega_a + \Delta$, and off-resonantly driving the $b$-mode at a detuning $\Delta$ and strength $\beta$~\cite{chamberland2022building} approximately realizes the Hamiltonian is given by:
\begin{equation}
    \hat{H} = \hbar g \hat{a}^\dagger \hat{b} + \mathrm{h.c.}, \label{eqn:bs}
\end{equation}
where the coupling rate $g$ is given by $\hbar g \equiv E_J\epsilon_p \beta \varphi_b^2\varphi_a$ thereby enabling the crucial swapping~\cite{makihara2024parametrically,chamberland2022building}. 

% Our transducer bandwidth will be limited by the largest feasible $g$. We note that to ignore strong driving from the first-order term in our Hamiltonian coming from $\sin\big(\hat{\varphi}_b + \hat{\varphi}_a)$, we require $2E_J\epsilon_p\varphi_b \simeq \hbar \omega_b \tau$ for small $\tau$. If we fix $\tau \simeq 1/5$ and assume $\varphi_b \simeq 0.5$ and $\omega_b/2\pi \simeq$ 5 GHz, this fixes $E_J\epsilon_p/ h \simeq$ 1 GHz. Further, adiabatic elimination of the $b$-mode is crucial for deriving the above effective Hamiltonian, so we require $\beta < 1$, although this constraint can be relaxed because spurious effects can be accounted for~\cite{chamberland2022building}. Finally, we estimate that $g_\mathrm{max}/2\pi \simeq$ 100 MHz should be possible, by diagonalizing the circuit with assumed values in design ?? (see Table ??). 

As discussed below, lower frequencies may lead to higher state-transfer fidelities and quantum channel capacities. The lowest frequency accessible by our transducer will largely be limited by how large of a capacitance can be fabricated. Capacitances larger than 1 pF will be challenging to fabricate in a fully planar way without spurious resonances, which would interfere with the higher frequency $b$-mode, and so parallel plate capacitors may be needed. Such capacitors are already compatible with superconducting qubit fabrication~\cite{weber2011single} and have been used to demonstrate capacitances exceeding 100 pF while maintaining $Q_\mathrm{i} > 10^5$~\cite{cicak2009vacuum,cicak2010low}. We show two possible designs in Table 1. % We assume that the intrinsic loss for the $a$-mode is dominated by direct coupling to the readout port. Similarly, we assume that the intrinsic loss for the $b$-mode is dominated by direct coupling to the NbTi cable, wherein we assume the NbTi cable is a continous bath. 
For both designs, we have confirmed that the direct coupling of the $b$-mode to the coaxial cable, which we model as a continuous bath, is less than $1$ kHz. Similarly we have confirmed that the direct coupling of the $a$-mode to the environment is less than $1$ kHz. We note that the transducer quality factors can be improved by coupling it to the readout with a high-pass filter and coupling it to the cable with a low-pass filter, both of which can be readily fabricated on the same chip.

Our transducer bandwidth will be limited by the largest feasible $g$. One upper bound on $g$ is the requirement that $g \ll \omega_a$ such that the $a$-mode is not too strongly coupled to the environment. For the two designs in Table 1, we believe that $g/2\pi\simeq$ 25 MHz should be possible. For example, applying a flux pump amplitude of $\epsilon_p/\pi = 0.03$ and an off-resonant drive on the buffer with amplitude $\beta = 6$ yields $g/2\pi = 25.20$ MHz for design A and $g/2\pi = 31.48$ MHz for design B~\cite{reglade2024quantum}.

Previous multi-octave coupling schemes based on longitudinal coupling~\cite{eichler2018realizing,bothner2021photon,rodrigues2021cooling} have a rate given by $g = g_0 \sqrt{n_c}$ where $g_0$ is the single-photon coupling rate and $n_c$ is the number of intracavity photons in the high-frequency resonator. These circuits have demonstrated up to $g_0/2\pi \simeq $ 175 kHz without adverse effects from flux sensitivity~\cite{rodrigues2024photon}, requiring $\gtrsim 20\times 10^3$ intracavity photons to achieve $g/2\pi = 25$ MHz. At such large intracavity photon numbers, even small Kerr shifts with approximate magnitudes of 1 kHz~\cite{bothner2021photon} can lead to large frequency shifts in the high-frequency resonator.  

% \begin{table}
% \caption{Transducer designs relating the circuit parameters (indicated in Figure 1a) to the Hamiltonian and input-output parameters.}
% \begin{ruledtabular}
% \begin{tabular}{cccccc}
% Design & $C_a$ & $C_b$ & $C_c$ \\
% A & 9.0 pF & 39.0 fF & 23.0 fF \\
% B & 1.0 pF & 30.0 fF & 16.0 fF\\
% \hline
% Design & $L_a$ & $L_b$ & $L_c$ \\
% A & 10.0 nH & 2.70$\mu$H & 70.0 nH \\
% B & 10.0 nH & 240.0 nH & 63.0 nH \\
% \hline
% Design & $\omega_a$ & $\omega_b$ & $\varphi_a$ & $\varphi_b$ \\
% A & 193.8 MHz & 6.427 GHz & 0.20 & 0.44 \\
% B & 690.8 MHz & 7.735 GHz & 0.32 & 0.46 \\
% \hline
% Design & $\kappa_a^e/2\pi$ & $\kappa_a^i/2\pi$ & $\kappa_b^e/2\pi$ & $\kappa_b^i/2\pi$ \\
% A & 110.8 MHz & 747.3 Hz & 109.9 MHz & 736.0 Hz \\
% B & 100.5 MHz & 34.8 kHz & 100.2 MHz & 23.3 kHz \\
% \end{tabular}
% \end{ruledtabular}
% \end{table}

\begin{table}
\caption{Transducer designs relating the circuit parameters (indicated in Figure 1a) to the Hamiltonian and input-output parameters.}
\begin{ruledtabular}
\begin{tabular}{cccccc}
Design & $C_a$ & $C_b$ & $C_c$ \\
A & 16.3 pF & 40.3 fF & 18.2 fF \\
B & 1.0 pF & 40.3 fF & 18.2 fF\\
\hline
Design & $L_a$ & $L_b$ & $L_c$ & $E_J/h$ \\
A & 45.7 nH & 23.3 nH & 71.9 nH & 3.51 GHz \\
B & 106.4 nH & 18.0 nH & 91.5 nH & 4.53 GHz \\
\hline
Design & $\omega_a$ & $\omega_b$ & $\varphi_a$ & $\varphi_b$ \\
A & 201.0 MHz & 5.302 GHz & 0.051 & 0.50 \\
B & 689.1 MHz & 5.304 GHz & 0.049 & 0.50 \\
\hline
Design & $\kappa_a^e/2\pi$ & $\kappa_b^e/2\pi$ \\
A & 50.0 MHz & 50.0 MHz \\
B & 50.0 MHz & 50.0 MHz \\
\end{tabular}
\end{ruledtabular}
\end{table}

\begin{figure}
\includegraphics[scale=1.05]{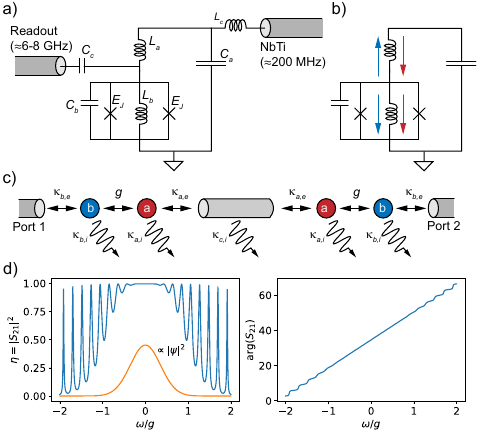}
\caption{(a) Schematic of the transducer circuit coupling the high-frequency readout port to the low-frequency modes of a NbTi cable. (b) Approximate illustration of the transducer's two modes. The blue (red) arrows indicate the high (low) frequency mode's current distribution through the inductors. (c) Illustration of the full system with a MOATS transducer on either end of a superconducting NbTi cable. (d) Example photon conversion efficiency for a cable with $Q_\mathrm{i} = 10^5$ and $L$ = 20 m and a transducer with coupling rates $\kappa_b^e = \kappa_a^e = 2g = 2\pi\times$50 MHz. The photon mode-shape $\psi$ is also shown in the inset, where the bandwidth is fixed to $g/2$.}
\label{f1}
\end{figure} 

%%%%%%%%%%%%%%%%%%%% FIDELITY AND ADDED NOISE %%%%%%%%%%%%%%%%%%%%
\section{Fidelity and added noise}
The fidelity and noise can be calculated from our system by considering the scattering parameters. Our full system is shown in Figure 1c, where we include intrinsic noise ports so we can inject thermal noise into our system. We model our cable as a 4-port lossy beamsplitter with efficiency $\eta_c$:
\begin{equation}
    S_{\mathrm{cable}} = 
    \begin{pmatrix}
        0 & \sqrt{\eta_{c}}e^{i\beta L} & 0 & \sqrt{1-\eta_{c}} \\
        \sqrt{\eta_{c}}e^{i\beta L} & 0 & \sqrt{1-\eta_{c}} & 0 \\
        0 & \sqrt{1-\eta_{c}} & 0 & -\sqrt{\eta_{c}}e^{i\beta L} \\
        \sqrt{1-\eta_{c}} & 0 & -\sqrt{\eta_{c}}e^{i\beta L} & 0
\end{pmatrix}
\end{equation}  
where ports 1 and 2 couple the two $a$-modes together, and port 3 and 4 allow for loss and added noise. The parameter $\eta_c= e^{-2\alpha L}$ is the power decay rate. We derive the scattering matrix for the transducer from input-output theory using the beam-splitter Hamiltonian in equation~(\ref{eqn:bs}). The total photon conversion efficiency $|\eta[\omega]|^2$ of the transducer is plotted in Figure 1d, assuming a constant cable-mode $Q_\mathrm{i} = 10^5$, $L =$ 20 m, and $\kappa_b^e = \kappa_a^e = 2g = 2\pi\times$50 MHz.

The most general way to characterize the channel is through the channel capacity, which we consider in the next section. Here we instead consider a simpler question: What is the state fidelity for single-photons after they are transmitted from one end of the system to the other? We assume photons with a Gaussian mode-profile for simplicity. Namely, our input operator is given by $\hat{A}_{in} = \sum_j \psi[\omega_j] \hat{a}_{in}[\omega_j]$ where $\psi[\omega_j]$ is a Gaussian in frequency domain and is normalized such that $\sum_j|\psi[\omega_j]|^2 = 1$ (as shown in Figure 1d). We've assumed here a discrete frequency axis with $\omega_{j+1}-\omega_j = \Delta \omega$. To calculate the fidelity, it is also helpful to calculate the added noise to our signal due to thermal states being injected into each noise port. We model this by injecting thermal states into the intrinsic loss channels of each $a$-mode, as well as the cable. We assume that the baths coupled to the higher frequency $b$-mode are in the vacuum state. The added noise from the $i^\text{th}$ noise port to the output (port 2) is given by:
\begin{equation}
    n_{\mathrm{added},i} = \sum_j |S_{2i}[\omega_j]|^2 |\psi[\omega_j]|^2 \bar{n}_{in,i}(\omega_j,T)
\end{equation}
where the total added noise is summed over all noise ports. 

Given the scattering parameters and total added noise, we can model our system as a Gaussian thermal loss channel with conversion efficiency $\eta$~\cite{holevo2001evaluating,weedbrook2012gaussian}. In the Heisenberg picture, we represent our channel by the transformation:
\begin{equation}
    \hat{A} \rightarrow \sqrt{\eta}\hat{A} + \sqrt{1-\eta}\hat{B},
\end{equation}
where
\begin{equation}
    \eta = \sum_j |S_{21}[\omega_j]|^2 |\psi[\omega_j]|^2
\end{equation}
is the conversion efficiency of our pulse. The effective input thermal noise is related to the total added noise by:
\begin{equation}
    n_{\text{added}} = (1-\eta) n_{\text{th}}
\end{equation}
where $n_{\text{th}} = \langle\hat{B}^\dagger\hat{B}\rangle$. Finally, the fidelity of our photon is given by $F = \langle 1 |\rho_s|1\rangle$ where $\rho_s$ is the final density matrix after tracing away the environment. 

Figure 2a and 2b respectively plot the fidelity and total added noise versus $a$-mode frequency at different cable temperatures. We assume a cable-mode $Q_\mathrm{i} = 10^5$, a cable length of $L = $100 m, an $a$-mode $Q_{\mathrm{i},a}=20\times10^3$, and a fixed $b$-mode intrinsic linewidth of $50$ kHz. We also assume the transducer is always thermalized to 10 mK, and that $\kappa_a^e = \kappa_b^e = 2g = 4BW = 2\pi\times$ 50 MHz, where $BW$ is the photon bandwidth (taken to be the standard deviation of $\psi[\omega]$), corresponding to a maximally flat transducer~\cite{wang2022quantum}. The vertical dotted lines in both figures indicate the frequencies of the proposed designs in Table 1, and the shaded regions indicate typical qubit frequencies ($\simeq$ 6-8 GHz). Figure 2a shows that, for a cable at 10 mK with a $Q_\text{i}=10^5$, the single photon fidelity at 200 MHz, 700 MHz, and 8 GHz is $\simeq$0.962, $\simeq$0.956, and $\simeq$0.772, respectively. This dramatic drop in fidelity reflects the increasing cable attenuation at higher frequencies. At higher temperatures, the improvement in cable loss is offset by added thermal noise. The number of added noise photons for a 10 mK cable at 200 MHz is $\simeq$ 0.0036 photons, compared to $\simeq$ 0.56 photons at 1 K. The dominant loss channel in this system is cable loss, where assuming a lossless transducer increases the fidelity of transduction at 200 MHz to $\simeq$ 0.966 and the fidelity of transduction at 8 GHz to $\simeq$ 0.787. However, using an identically lossy transducer and increasing the cable-mode $Q_\mathrm{i}$ to $10^6$ increases the single photon fidelity at 200 MHz, 700 MHz, and 8 GHz is $\simeq$0.976, $\simeq$0.975, and $\simeq$0.941, respectively. Figure 2c plots the fidelity versus $L/Q_\mathrm{i}$ for a cable thermalized to 10 mK, where we sweep $Q_\mathrm{i}$. Figure 2c shows that for a cable-mode $Q_\mathrm{i} = 10^6$, as has recently been demonstrated in Al coaxial cables~\cite{niu2023low}, and assuming a cable temperature of 10 mK, single photons can propagate for 1 km while maintaining fidelities above 0.962 if they are transduced down to $200$ MHz, but that fidelity drops to 0.772 if they are transmitted at $8$ GHz. % Alternatively, it shows that for a fixed cable $Q_\mathrm{i}$, a single-photon transduced to 200 MHz can propagate over sixty times further than one transduced to 8 GHz while maintaining the same fidelity.

\begin{figure}
\includegraphics[scale=1.05]{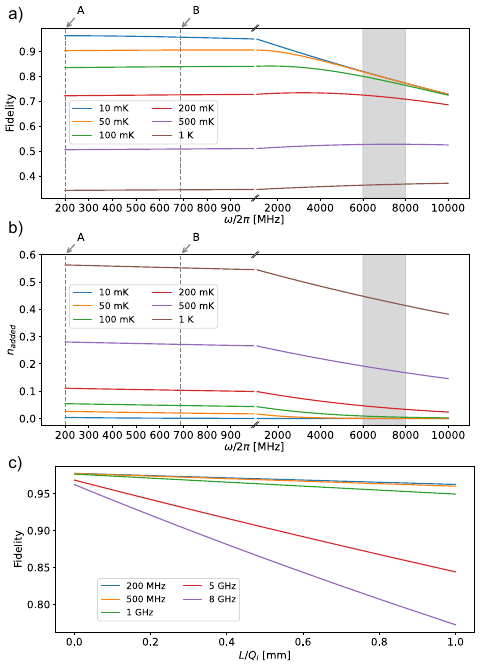}
\caption{(a) Single-photon fidelity versus $a$-mode frequency for different cable temperatures. (b) Total added noise versus $a$-mode frequency for different cable temperatures. The cable $Q_\mathrm{i}= 10^5$ and $L =$ 100 m. We assume the transducer is always thermalized to 10 mK, and that $\kappa_a^e = \kappa_b^e = 2g = 4BW = 2\pi\times$ 50 MHz, where $BW$ is the photon bandwidth. We assume the $a$-mode $Q_\mathrm{i}=20\times10^3$ and that $\kappa_b^i =2\pi\times 50$ kHz. The dotted vertical lines indicate the frequencies of the designs presented in Table 1, and the shaded region indicates typical qubit frequencies ($\simeq$ 6-8 GHz). (c) Single-photon fidelity versus $L/Q_\mathrm{i}$.}
\label{f1}
\end{figure} 

%%%%%%%%%%%%%%%%%%%% QUANTUM CHANNEL CAPACITY %%%%%%%%%%%%%%%%%%%%
\section{Quantum channel capacity}
A more general understanding of our proposed approach's capability requires us to estimate the maximum rate at which qubits can be transmitted across the imperfect channel formed by the transducers and coaxial cable, and compare this to a direct transmission approach. Unlike the single-photon fidelity, this figure of merit is an intrinsic property of the noisy channel and is independent of the input state or its spectrum. We consider the \textit{continuous-time one-way thermal loss capacity}~\cite{wang2022quantum}, which we will simply refer to as the channel capacity. This definition takes into account the finite conversion band of our transducer by considering different frequency modes as parallel quantum channels. Currently, there is no analytic expression for the channel capacity of a general Gaussian thermal-loss channel, but there are well-known upper and lower bounds. The lower (upper) bound on channel capacity is given by:
\begin{equation}
	Q_{L(U)} = \int q_{L(U)}[\omega]\frac{d\omega}{2\pi}
\end{equation}
where $q_{L(U)}$ is the lower (upper) bound on the so-called discrete-time thermal-loss capacity. The best known lower bound on $q[\omega]$ is given by~\cite{holevo2001evaluating}:
\begin{equation}
	q_{L}[\omega] = \max\Big\{\log_2\Big[\frac{\eta[\omega]}{1-\eta[\omega]}\Big],0\Big\}
\end{equation}
We consider the upper bound on $q[\omega]$ that is obtained from a twisted version of the attenuator decomposition of thermal attenuators~\cite{rosati2018narrow,noh2018quantum}:
\begin{equation}
	q_{U}[\omega] = \max\Big\{\log_2\Big[\frac{\eta[\omega]-(1-\eta[\omega])\bar{n}[\omega]}{(1-\eta[\omega])(\bar{n}[\omega]+1)}\Big],0\Big\}
\end{equation}
We found that this upper bound was tighter than the bound derived from degradable extensions of thermal-loss channels for our parameter regime~\cite{fanizza2021estimating}.

Figure 3a plots the upper (dashed line) and lower (solid line) on quantum channel capacity (in units of the parametric $g$ coupling the $a$-mode and $b$-mode) versus $a$-mode frequency at different cable temperatures. The assumed cable and transducer parameters are identical to those in Figure 2a and 2b. The vertical dotted lines in both figures indicate the frequencies of the proposed designs in Table 1. Figure 3a shows that, for a cable at 10 mK with $Q_\text{i} = 10^5$, the lower bound on channel capacity at 200 MHz is $\simeq 1.43\times g$ whereas at 8 GHz it is $\simeq 0.48\times g$. This reflects the improvement in attenuation at lower frequencies. The dominant loss channel is the cable loss, where assuming a lossless transducer increases the lower bound on capacities at 200 MHz and 8 GHz to $\simeq 1.48\times g$ and $\simeq 0.52\times g$, respectively. Figure 3b plots the channel capacity versus $L/Q_\mathrm{i}$ at different frequencies for a cable thermalized to 10 mK, where we seep $Q_\mathrm{i}$. % Figure 3b shows that, for a fixed cable $Q_\mathrm{i}$, information can propagate over sixty times further when transduced to 200 MHz as compared to 8 GHz while maintaining the same channel capacity.

\begin{figure}
\includegraphics[scale=1.05]{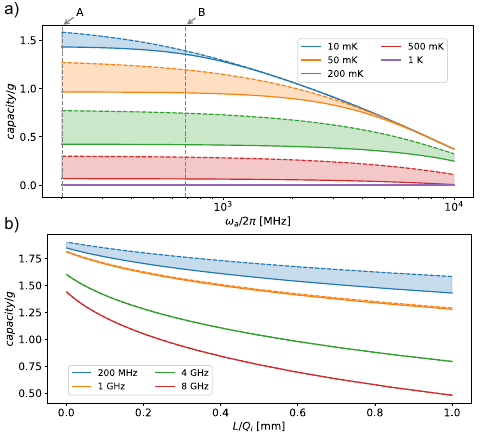}
\caption{Upper bounds (dashed lines) and lower bounds (solid lines) on quantum channel capacity versus $a$-mode frequency for different cable temperatures (a) and versus $L/Q_\mathrm{i}$ for a 10 mK thermalized cable for different transduction frequencies (b). The assumed cable and transducer parameters are identical to those presented in Figure 2a and 2b. The dotted vertical lines in (a) indicate the frequencies of the designs presented in Table 1.}
\label{f1}
\end{figure} 

%%%%%%%%%%%%%%%%%%%% CONCLUSION %%%%%%%%%%%%%%%%%%%%
\section{Conclusion}
We proposed networking superconducting quantum circuits through cryogenic links using superconducting coaxial cables as opposed to microwave waveguides. We showed that accessing the low frequency cable-modes provides a dramatic increase in single-photon fidelity and quantum channel capacity, and we proposed accessing these modes using a modified circuit based on an ATS. Our proposal provides a simple, highly efficient, large bandwidth approach to transduction. In addition to figures of merit based on transduction, superconducting coaxial cables provide other benefits over microwave waveguides, such as flexibility and a much smaller footprint. 

%%%%%%%%%%%%%%%%%%%% ACKNOWLEDGEMENTS %%%%%%%%%%%%%%%%%%%%
\section{Acknowledgements}
We thank Liang Jiang, Nathan Lee, Yudan Guo, and Zhaoyou Wang for useful discussions.  T.M. acknowledges support from the National Science Foundation Graduate Research Fellowship Program (grant no. DGE-1656518). A.H.S.-N. acknowledges support via a Sloan Fellowship.  The authors also wish to thank NTT Research and Amazon Web Services Inc. for their financial support. Some of this work was funded by the U.S. Department of Energy through Grant No. DE-AC02-76SF00515 and via the Q-NEXT Center. This material is based upon work supported by the Air Force Office of Scientific Research and the Office of Naval Research under award number FA9550-23-1-0338. Any opinions, findings, and conclusions or recommendations expressed in this material are those of the author(s) and do not necessarily reflect the views of the United States Air Force or the Office of Naval Research.

\newpage
\bibliography{references.bib} % Entries are in the refs.bib file

\end{document}